\documentclass[10pt,doubleside]{IEEEtran}
\usepackage{pifont,multicol,amsfonts,amsmath,color,amssymb,graphicx,tabularx,epsfig, cite,psfrag,subfigure,algorithm,enumerate,stfloats,algorithmic,epstopdf,balance}

\begin{document}
\title{AIRIS: Artificial Intelligence Enhanced Signal Processing in Reconfigurable Intelligent Surface Communications} 	
\author{Shun Zhang, \emph{Senior Member, IEEE}, Muye Li, Mengnan Jian, Yajun Zhao, Feifei Gao, \emph{Fellow, IEEE}

\thanks{S. Zhang and M. Li are with the State Key Laboratory of Integrated Services Networks, Xidian University, Xi'an 710071, P. R. China (e-mail: zhangshunsdu@xidian.edu.cn; myli$\_$96@stu.xidian.edu.cn).}
\thanks{M. Jian and Y. Zhao are with Algorithm Department of Wireless Product R$\&$D Institute, ZTE Corporation, Beijing 100029, China (e-mail: jian.mengnan@zte.com.cn; zhao.yajun1@zte.com.cn).}
\thanks{F. Gao is with Institute for Artificial Intelligence, Tsinghua University (THUAI), State Key Lab of Intelligent Technologies and Systems, Tsinghua University, Beijing National Research Center for Information Science and Technology (BNRist), and Department of Automation, Tsinghua University Beijing, 100084, P.R. China (e-mail: feifeigao@ieee.org).}
}
\maketitle

\begin{abstract}
Reconfigurable intelligent surface (RIS) is an emerging meta-surface that can provide additional communications links through reflecting the signals, and has been recognized as a strong candidate of 6G mobile communications systems. Meanwhile, it has been recently admitted that implementing artificial intelligence (AI) into RIS communications will extensively benefit the reconfiguration capacity and enhance the robustness to complicated transmission environments.
Besides the conventional model-driven approaches, AI can also deal with the existing signal processing problems in a data-driven manner via digging the inherent characteristic from the real data. Hence, AI is particularly suitable for the signal processing problems over RIS networks under unideal scenarios like modeling mismatching, insufficient resource, hardware impairment, as well as dynamical transmissions.
As one of the earliest survey papers, we will introduce the merging of AI and RIS, called AIRIS, over various signal processing topics, including environmental sensing, channel acquisition, beamforming design, and resource scheduling, etc.
We will also discuss the challenges of AIRIS and present some interesting future directions.

\end{abstract}

\maketitle
\thispagestyle{empty}


\section{Introduction}
Standing at the beginning of the fifth-generation (5G) era, the communication academia and industry have started the research procedure of the next-generation communications system  (6G), in which the typical frameworks of 5G, i.e., the enhanced mobile broadband (eMBB), the ultra-reliable \& low-latency communication (URLLC), and the massive machine-type communications (mMTC), demand for further enhancement.
To cover all these requirements, a variety of techniques booms up. For example, a passive meta-surface called the reconfigurable intelligent surface (RIS) has drawn the extensive attention  \cite{metasurface}, whose general application is to fight against the blockage of the line-of-sight (LOS) path and improve the communication quality through effectively reconfiguring the propagation of electromagnetic (EM) signal.
Especially, since the propagation paths at mmWave frequency band are very sensitive to the blockages in the environment \cite{mmwave2,mmwave3,mmwave4}, RIS can perfectly match the requirement of mmWave communications~\cite{mmwave1}.
Different from the relay or backscatter communications,
RIS only reflects the impinging signal in a reconfigurable way and does not need the complicated  signal processing units, such as the power amplifiers, the mixers, and the baseband units \cite{signal_processing_units}.
Moreover,  the reflecting direction is controllable by adjusting the phase shift and/or the amplitude of each element on RIS, and thus  a desired beam can be formulated towards the objective user \cite{Rui_Zhang_Magazine}.
Thus, RIS communications has lower power consumption and manufacturing cost, and has been recognized as a strong candidate  technique of 6G.

Enormous  signal processing techniques   were utilized to deals with the communications issues in RIS \cite{sensing_1, traditional_1, traditional_3, traditional_4}, such as the sparse Bayesian learning, the iterative optimization, and the orthogonal matching pursuit algorithms.
However,  conventional signal processing approaches exhibit certain drawbacks.
On the one hand, conventional signal processing techniques are closely dependent on hypothetical mathematical models.
In practical scenarios, the rapid changing of the radio scattering and the hardware impairments will cause serious mismatch and nonlinear characteristics to the assumed mathematical models, which will deteriorate the performance of  conventional signal processing techniques \cite{J_Ma_model_mismatching}.
On the other hand, conventional optimization techniques in complicated transmission scenario will converge to local optima and face high computational complexity with large meta surface.

Artificial intelligence (AI) is well known for its capability to map nonlinear models caused by any complicated scenario, and has been widely applied  in the areas of image recognition, speech recognition, text recognition, computer vision, etc.
For a non-model situation, AI could approximate unknown relations from sufficient real data \cite{survey_DL_RIS}.
Naturally, re-investigation of the signal processing problems  of RIS communications, e.g., the environmental sensing, the channel acquisition, the beamforming design, and the resource scheduling, via AI techniques has become a major research topic.

\begin{table*}[htbp]
\centering
\caption{Different AI algorithms with distinguished RIS scenarios.}\label{table_1}
\begin{scriptsize}
\begin{tabular}{|p{5.1cm}|p{10.9cm}|}
  \hline \hline
  \bf{Type of AI algorithms} & \bf{Application scenarios}  \\
  \hline
  \multicolumn{2}{|c|}{\bf{Environmental sensing}} \\
  \hline
  Supervised learning \cite{Fingerprinting} & Localization problem in the system with only one AP and one RIS \\
  \hline
  Machine learning \cite{imaging} & Image-based sensing with a high resolution image of the propagation environment \\
  \hline
  \multicolumn{2}{|c|}{\bf{Channel acquisition}} \\
  \hline
  Deep residual learning \cite{DL_channel_est_1} & Direct cascaded channel estimation over the  RIS communications \\
  \hline
  Federated learning \cite{Federated_Learning_estimation} &  Channel estimation in multi-user case\\
  \hline
  Supervised learning \cite{Meng_Xu_RIS} & Antenna domain cascaded channel extrapolation with the full passive RIS \\
  \hline
  Supervised learning \cite{Shunbo_extrapolation} & Antenna domain individual channel extrapolation with the hybrid active/passive RIS \\
  \hline
  \multicolumn{2}{|c|}{\bf{Beamforming design}} \\
  \hline
  Supervised learning \cite{Chongwen_Huang_Indoor} & Beamforming design with passive RIS under the indoor scenario \\
  \hline
  Deep reinforcement learning \cite{Chongwen_JSAC} & Beamforming design with the learning of the environment \\
  \hline
  Supervised learning \cite{Precoding_Dai} & Solving the non-convex problem to maximize the system sum-rate in the RIS based hybrid precoding system \\
  \hline
  Supervised learning \cite{Shunbo_extrapolation} & Simultaneously optimizing the  RIS activation pattern and the RIS beamforming
   in the hybrid passive/active RIS system \\
  \hline
  Supervised learning \cite{Drones} & Beam tracking under the high mobility scenario \\
  \hline
  \multicolumn{2}{|c|}{\bf{Resource scheduling}} \\
  \hline
  Deep reinforcement learning \cite{RIS_NOMA1} & User partitioning and RIS beamforming for the RIS assisted  NOMA networks \\
  \hline
  Supervised learning \cite{mmwave11} & Improving the performance of RIS assisted mmWave communications \\
  \hline
  Deep reinforcement learning \cite{THz_DRL} & Joint scheduling of multiple RISs to enlarge the sensing range of the RIS assisted communications \\
  \hline
  Deep reinforcement learning \cite{Scheduling_DRL} & Maximizing the average energy efficiency through joint optimizing the power allocation, RIS beamforming, and RIS elements' ON/OFF state \\
  \hline
  Federated learning \cite{mmwave_FL} & Privacy-preserving design in the RIS assisted mmWave system \\
  \hline
  Federated learning \cite{FL_multiRIS} & Minimizing the aggregation error and accelerating the convergence rate under the static scenario with multiple RIS \\
  \hline
  Multi-agent deep reinforcement learning \cite{Scheduling_Timevarying_Multicell} & Dynamic control of the RISs under the  mobility scenario \\
  \hline
  Deep reinforcement learning \cite{vehicle_comm} & Joint terminal scheduling and passive beamforming over the RIS empowered vehicular communications \\
  \hline
  \hline
\end{tabular}
\end{scriptsize}
\end{table*}

Most existing survey papers of RIS communications mainly focus on the conventional signal processing aspects while few work has summarized its integration with AI.
Without any applications of AI, the RIS is not real ``intelligent surface''.
To fill in this blank, we here present a survey to the merging of AI and RIS communications (AIRIS) and demonstrate the resultant improvement compared with traditional techniques.
To this end,
we will introduce AI assisted environmental sensing, AI assisted channel acquisition, AI assisted beamforming design, and AI assisted resource scheduling in different scenarios, as shown in Table~{\ref{table_1}}.
Except for the state-of-the-art researches on AIRIS, we will also present some future research directions aiming to provide diverse technical discussions and facilitate the continuous research on AIRIS.

\vspace{-0mm}
\section{Key Issues of RIS Communications}
\subsection{Signal Processing Problems}
To enhance the performance of RIS communications, several critical signal processing issues should be addressed: the environmental sensing, the channel acquisition, the beamforming design, and the resource scheduling \cite{RIS_signal_processing_problems}.
Environmental sensing is to get the localizations or the images of the objectives with the aid of the RIS. The channel acquisition
is to either obtain the cascaded channels or the individual channels at both sides of the RIS, which is more complicated than those in conventional MIMO systems. Note that different channel state information (CSI) will result into different strategies in the subsequent beamforming design. The beamforming design is a  key issue, where the phase shifter of the RIS should be well controlled to adjust the reflected beam in the desired direction.
Moreover, the resource scheduling includes the joint scheduling for the beamforming of multiple base stations (BSs) and multiple RISs, the users allocation, and the BS power allocations, which can help construct an optimal network topology and improve the system efficiency.
For example, the user-RIS pairs, or even the user-RIS-BS links, can be well scheduled, such that the users at close positions can be separately served by different BSs.

\subsection{Key Ideas of AIRIS}
\begin{figure*}[htbp]
 \centering
 \includegraphics[width=6.7in]{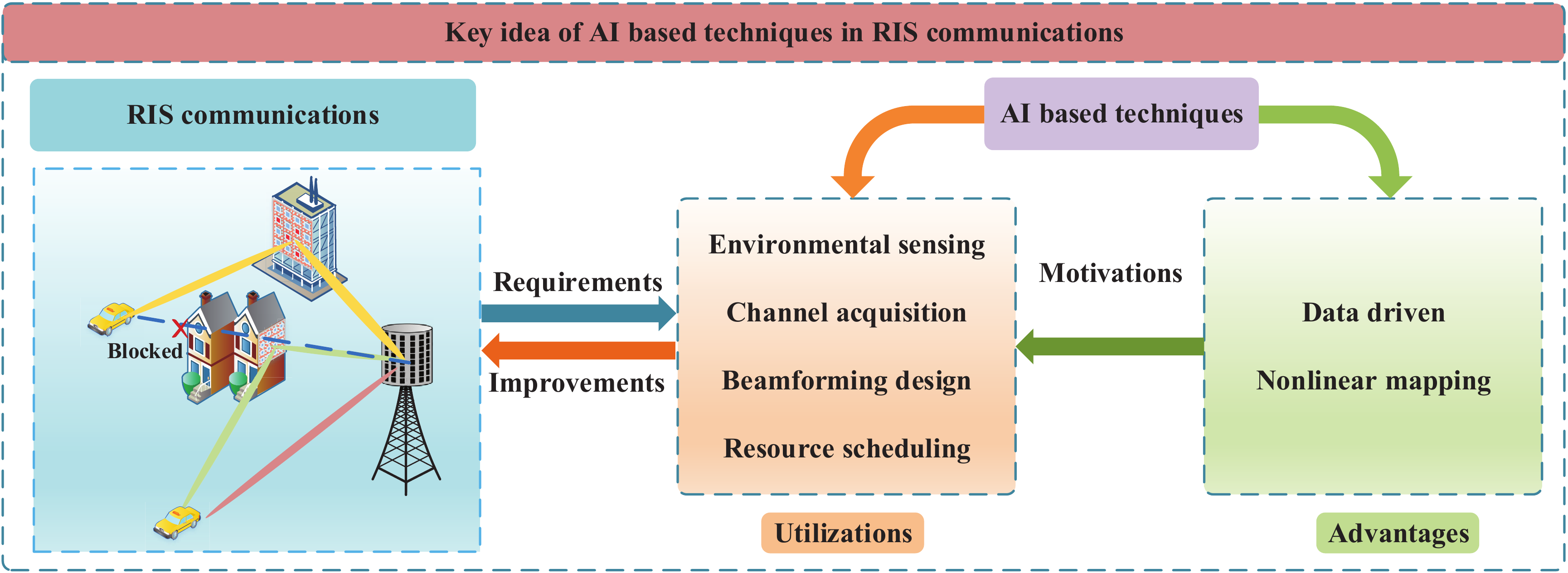}
 \caption{Key idea of AIRIS.}
 \label{fig:1}
\end{figure*}

The key idea of AIRIS is to solve the critical signal processing issues in RIS communications with AI based techniques, so to obtain higher system efficiency comparing to conventional techniques, as illustrated in \figurename{ \ref{fig:1}}.
Through training with huge dataset, the environmental sensing and channel acquisition can be more accurate.
Moreover, It can also help overcome the nonlinear problem caused by hardware impairments and pilot contamination.
Besides, with diverse learning strategy of AI, the beamforming design can be configured for different purposes, and the resource scheduling can also be implemented to construct a better network topology.
Hence, for the beamforming design and resource scheduling, AI will help enhance the robustness of the overall design.

\section{AI Assisted Sensing and Channel Acquisition}
\begin{figure*}[htbp]
 \centering
 \includegraphics[width=6.6in]{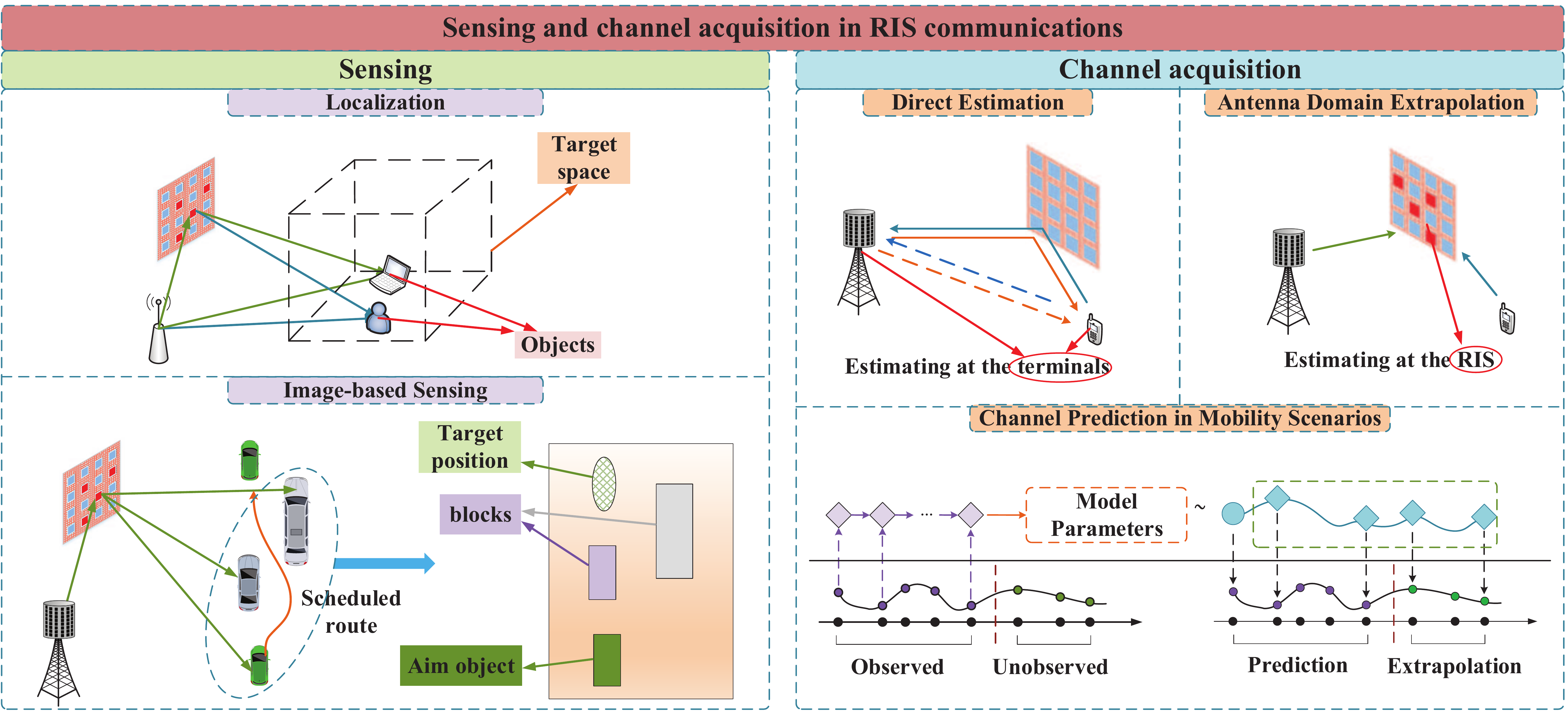}
 \caption{Scenarios of environmental sensing and channel acquisition for RIS communications.}
 \label{fig:2}
\end{figure*}

For RIS communications, the signal from the transmitter sequentially goes through
the link from the transmitter to RIS and that between RIS and the user.
Similar to the relay networks, there exists two types of channels, the cascaded channel and the individual ones.
The typical scenarios for the AI assisted environmental sensing and channel acquisition are illustrated in \figurename{ \ref{fig:2}}.

\subsection{AI Assisted Environmental Sensing}
\subsubsection{Localization}
\begin{figure}[htbp]
 \centering
 \includegraphics[width=3.2in]{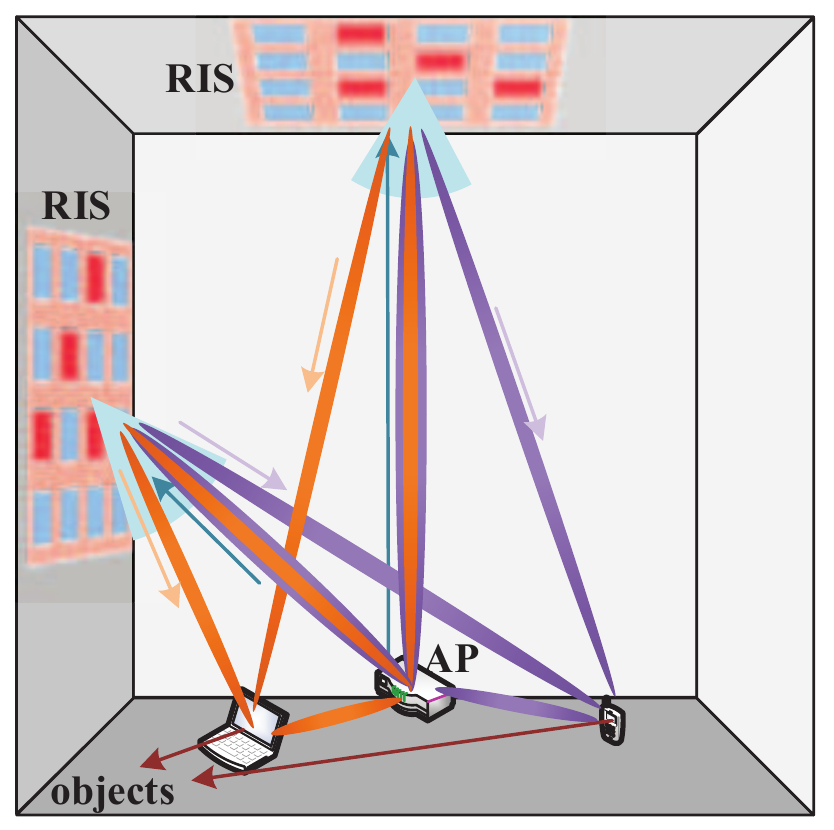}
 \caption{Illustration of environmental sensing with RISs.}
 \label{fig:Localization}
\end{figure}

In typical wireless fingerprint localization problem, one will adopt multiple access points (APs), and each AP can contribute towards one fingerprint.
If only one AP can be adopted, then different fingerprints can still be obtained with the aid of one RIS by shifting the feature of the RIS in different ways, i.e., reconfiguring the beamforming matrix of the RIS, as shown in \figurename{ \ref{fig:Localization}}.
Although a longer fingerprint vector can generally improve the localization accuracy via  signal mapping differentiability (SMD) and robustness, its creation will lead to large time consumption.
Hence, it is required to select a much smaller number of RIS features while simultaneously maintain the SMD to ensure the localization accuracy.
Since selecting optimal RIS features from the huge set of all possible candidates is very difficult, the authors of \cite{Fingerprinting} applies supervised learning based scheme for the RIS feature selection, which can reduce the time cost for position acquisition and simultaneously enhance the localization accuracy.

\subsubsection{Image-based Sensing}
In \cite{imaging}, an image-based sensing technique is proposed to fully exploit the received signal power at each RIS element, where
the power samples are processed to generate a high resolution image of the propagation environment.
Besides, the authors resort to the transfer learning and the support vector machine,  and design a computer vision algorithm to detect anomalies over a predefined route.

\subsection{AI Assisted Channel Acquisition}
\subsubsection{Direct Channel Estimation}

In \cite{DL_channel_est_1}, the authors propose to directly estimate the cascaded channels of RIS communications and formulate the channel estimation problem as a denoising process.
Then a deep residual learning (DReL) approach is adopted to implicitly learn the residual noise for recovering the channel coefficients from the noisy pilot-based observations.
Next, a deep residual network-based minimum mean square error (MMSE) estimator is derived with the aid of Bayesian philosophy.
However, in multi-user case, since the channel training of all the users are always centrally implemented at the BS, all the training dataset should be transmitted to the BS. Thus, its transmission overhead will be very high.
In \cite{Federated_Learning_estimation}, the authors adopt the federated learning (FL) framework, in which the neural networks will be trained at the users, rather than the BS.
In this case, only the updated parameters are required to be mutually transmitted.
As has been demonstrated in \cite{Federated_Learning_estimation}, the FL based approach enjoys approximately 16 times less transmission overhead compared to the centralized learning-based training, while performs in a similar way.

\subsubsection{Antenna Domain Extrapolation}
Due to the large scale of both the BS antenna arrays and the RIS elements, the direct estimation of the RIS channel will lead to huge training overhead.
One solution is to reconstruct the channel matrix from limited measurements through compressive sensing technique. However, compressive sensing approach needs complex mathematical operations and is not robust to the noisy measurements.

It was recently shown that the full channel can be  extrapolated from the partial one by AI assisted approach.
For full passive RIS, the extrapolation of the full cascaded channel can be implemented at the terminals, i.e., the BSs or the users.
In \cite{Meng_Xu_RIS}, the authors adopt the ordinary differential equation (ODE) and set up connections between different data layers in a convolutional neural network (CNN), which can approximate the mapping between sub-sampled channel and the full channel better than common CNNs. Meanwhile, in a recently designed hybrid active/passive RIS architecture, some RIS elements are enabled with signal processing abilities \cite{hybrid_passive_active}.
Thus the estimation of the individual channels can be directly implemented at the RIS \cite{Ahmed_enabling}.
In \cite{Shunbo_extrapolation}, the authors activate a fraction of RIS elements, recover the individual channels at the activated elements,
and then extrapolate the full individual channel information from the partial ones, where a CNN is adopted.
Moreover, the authors propose an antenna selection network based on the probabilistic sampling theory to select the optimal locations of those active RIS elements.
Both  \cite{Meng_Xu_RIS} and \cite{Shunbo_extrapolation} have demonstrated great effectiveness and strong robustness.

Recently, the element-grouping strategy is proposed to decrease the cascaded channel estimation overhead \cite{ele_grouping1}.
In this scheme, every RIS element in each group stays in the open state, shares the same reflection coefficient and is assumed to have the same CSI \cite{ele_grouping2}.
Actually, the channels within one element-group are not the same.
Only partial but not the entire channel information can be achieved in the element-grouping scheme.
Moreover, the achieved channels related with different RIS elements within each group would interfere with each other.
Inspired by the antenna extrapolation, one can design an NN to eliminate the interference within each element group and acquire the refined partial cascaded channels.
With these partial channels, one can further utilize another DL scheme to extrapolate the full cascaded channels.

\subsubsection{Channel Prediction in High Mobility Scenarios}
The channel estimation over RIS communications should also embrace
the high mobility scenario, in which the antenna domain channel extrapolation   \cite{Shunbo_extrapolation} can be transplanted into  the time domain.
It can be checked that the channel extrapolation over the time domain is essentially the time series reconstruction and prediction problem, for which the recurrent neural network (RNN) can be applied.
However, when the time-varying RIS channels possess the  long-term dependencies, the RNN model will face difficulties in gradient vanishing or exploding.
An alternative solution is to design a long short-term memory (LSTM) based scheme, which separates its memory from the time-continuous state.
However, due to the hardware impairment and environment interferences, part of the regularly sampled points may lose their efficiency, and only irregularly sampled channel can be obtained.
Inspired by the ODE structure, the continuous time-varying channel model can be constructed by adding connections between the irregularly sampled points with coefficients and linear calculations.
In this situation, the recently proposed latent-ODE can be implemented to obtain supreme  channel extrapolation performance over the time domain  \cite{latent_ODE}.
Different from traditional ODE model, in the latent-ODE, the latent variable framework can explicitly decouple the variation of the system, the likelihood of observations, and the recognition model, further allowing each to be examined or specified on its own \cite{latent_ODE_0}.
Hence, it will become easier to handle time domain prediction problem based on randomly sub-sampled measurements.
Compared with RNN and conventional ODE, the latent-ODE based channel prediction shows significant improvements on its predictive MSE.

\subsubsection{Integration of Different NNs}
To implement the channel extrapolation in either the antenna domain or the time domain, one of the important things is the accurate acquisition of the initial inputs, i.e., the estimation of sub-sampled channel.
Take the antenna domain channel extrapolation as an example, the sub-sampled channel can be obtained by the direct channel estimation techniques.
However, the estimation error of the sub-sampled channel will be brought into the later antenna selection and channel extrapolation network.
Passing through the NNs, the error will be amplified, and will further harm the channel extrapolation performance.

Two approaches can be used to solve this problem. The first one is to add some structures between the sub-sampled channel estimation process and the channel extrapolation one, e.g., the denoising network.
After training the denoising network, the estimation error can be learned and removed.
This idea can also be applied for the time domain extrapolation over the high mobility scenarios.
The second approach is to exquisitely design the original extrapolation network in order to deal with error-contaminated inputs.
This method can simplify the whole AI structure for the channel extrapolation and reduce the message passing complexity among the NNs.
\begin{figure*}[htbp]
 \centering
 \includegraphics[width=6.6in]{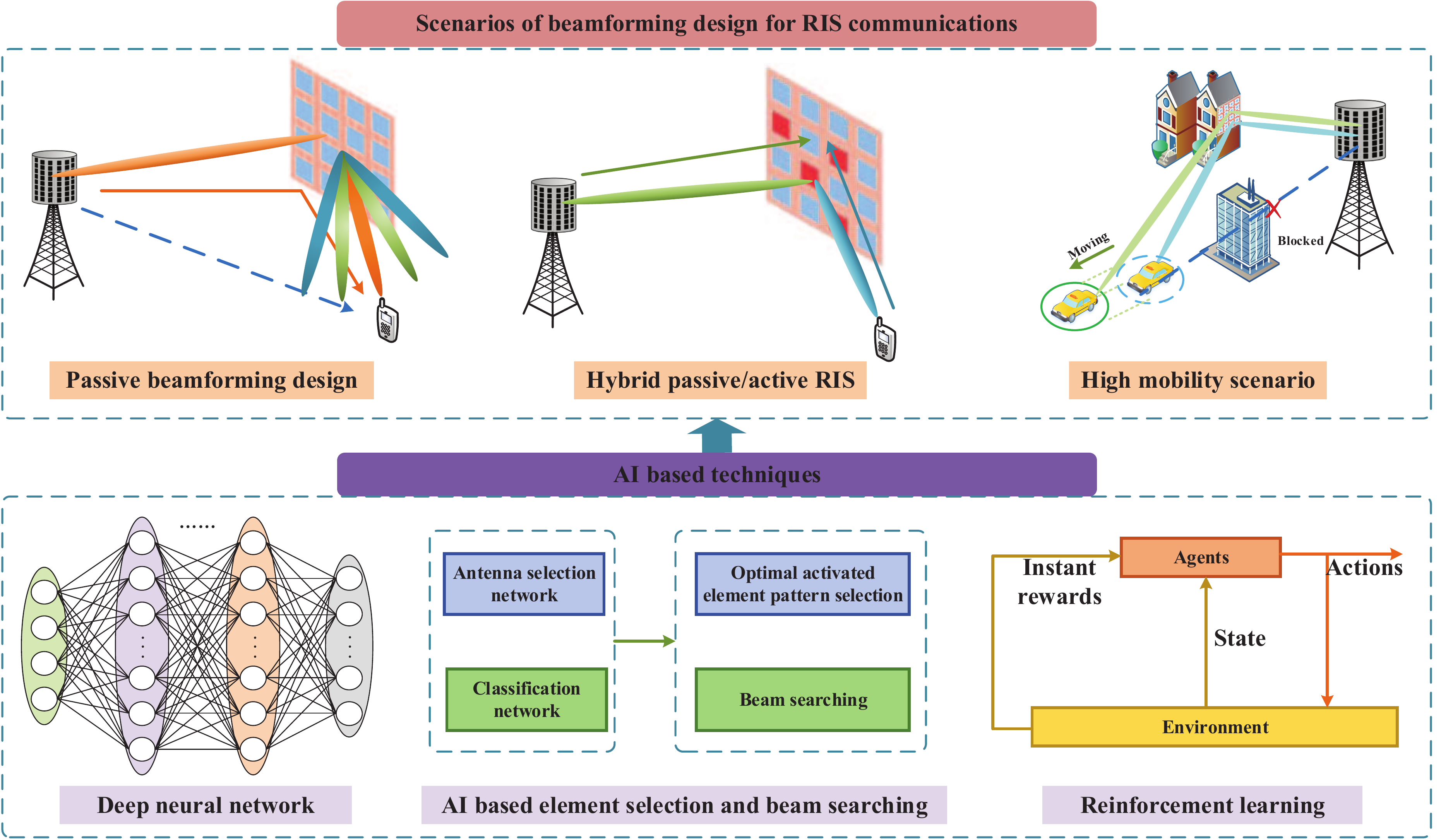}
 \caption{AI based beamforming design for RIS communications.}
 \label{fig:3}
\end{figure*}

\section{AI Assisted Beamforming}

The most important advantage of RIS communications comes from the beamforming that can control the signal reflection towards the desired direction and improve the quality of the transmission.
This can be achieved by continuously tuning the phase shift and the reflecting amplitude of the RIS elements.
However, high-precision adjustment will result into significant hardware cost  and huge challenge in the realization of control circuit.
Hence, implementing a discrete adjustable phase shift and amplitude for RIS is more appropriate \cite{Rui_Zhang_Magazine}. Here, we will present  typical AI based beamforming design strategies over three kinds of scenarios, i.e., the fully passive RIS, the hybrid passive/active RIS, and the high mobility cases, as shown in \figurename{ \ref{fig:3}}.

\subsection{Beamforming Design with Passive RIS}
For the indoor communication scenario, \cite{Chongwen_Huang_Indoor} presents a DL based method to implement the RIS configuration, where the deep neural network (DNN) is trained to accurately map the user's location and the configuration of the RIS's elements.
Then the optimal RIS beamforming matrix can be acquired through maximizing the received signal strength at the user side.
However, this supervised learning based approach needs large datasets and requires a long training time.
In \cite{Chongwen_JSAC}, the authors develop a deep reinforcement learning (DRL) based algorithm, where one agent will gradually derive its best action through the trial-and-error interactions with the environment over time, and then yield  the optimal beamforming matrices.
This algorithm can not only learn from the environment and gradually adjust its behaviour, but also obtains the comparable performance with the weighted MMSE and the zero forcing beamforming based fractional programming.

There are also some works about the RIS based hybrid precoding architecture, where the phased array based analog precoding at BS is replaced by the RIS beamforming \cite{Hybrid1,Hybrid2,Hybrid3}.
With the non-convex constraint on the discrete phase shifts, the problem of maximizing the system sum-rate is difficult to solve.
In \cite{Precoding_Dai}, the authors formulate a parallel DNN based classification problem and propose a DL based multiple discrete classification (DL-MDC) hybrid precoding scheme.
In this structure, multiple DNNs are utilized, and the output of each DNN corresponds to a diagonal element of the RIS based analog beamforming matrix.
Compared with the cross entropy optimization \cite{CEO_alg_in_Precoding_Dai}, the DL-MDC based scheme can reduce the runtime significantly with a negligible performance loss, and works well in both  the Saleh-Valenzuela channel model and the practical 3GPP one.

\begin{figure*}[htbp]
 \centering
 \includegraphics[width=6in]{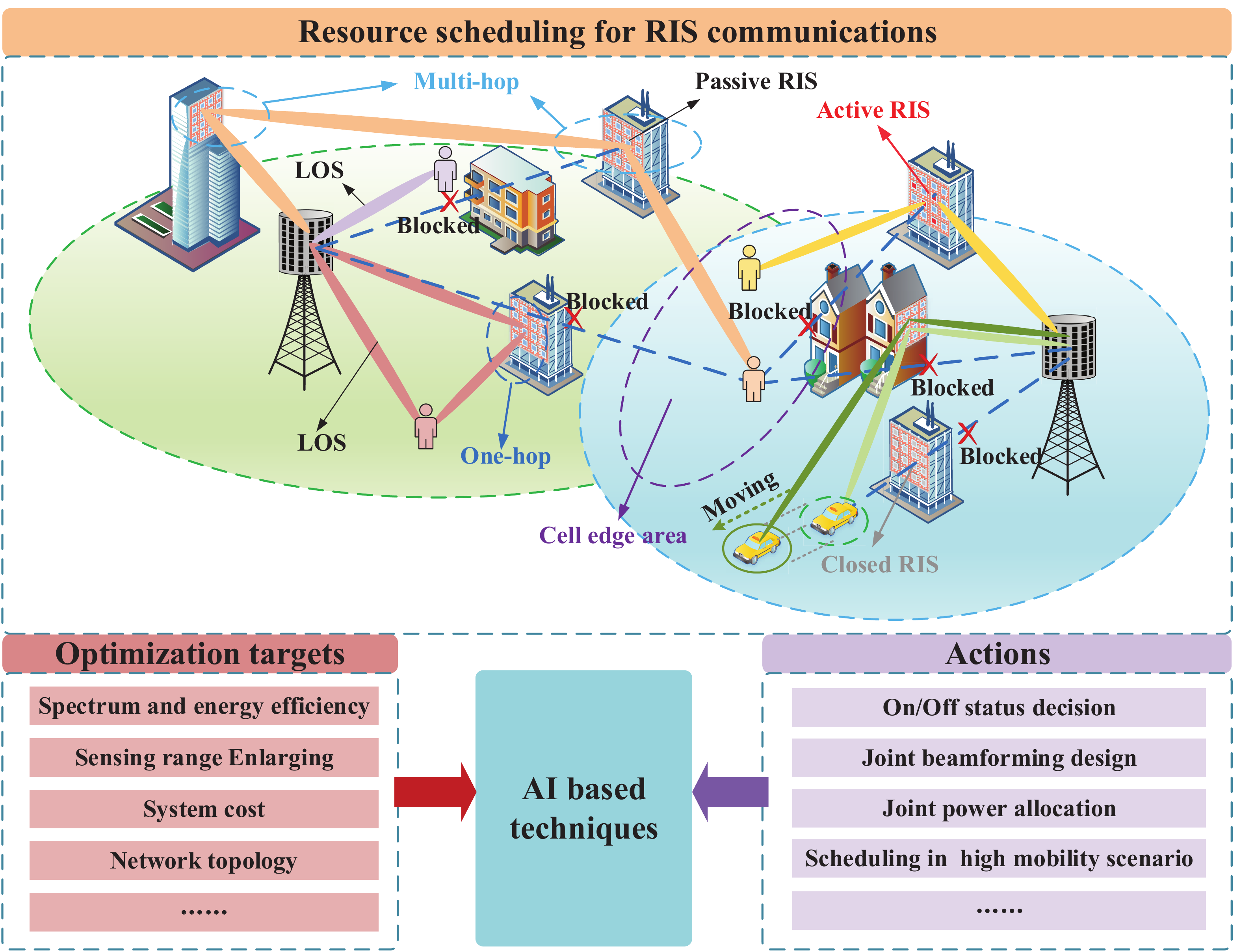}
 \caption{AI based resource scheduling for RIS communications.}
 \label{fig:4}
\end{figure*}

\subsection{Beamforming Design with Hybrid Passive/Active RIS}
For hybrid passive/active case, the RIS can only obtain and process sub-sampled channel information from the activated elements.
Compared with passive beamforming design,
since the RIS can only acquire effective signal from limited activated reflective elements, the input data of the NNs for the beamforming design will be limited,
which will harm the performance of the continuously optimization based scheme in \cite{Chongwen_JSAC}.
In \cite{Shunbo_extrapolation}, the authors construct a finite beamforming codebook with the aid of the DFT matrix, and the optimal beamforming selection from the codebook can be regarded as a classification problem.
Then, the authors design a fully-connected neural network (FNN) based classification network to jointly obtain the optimal activated reflection element pattern and the searched beamforming matrix of the hybrid passive/active RIS. Interestingly, with the extrapolated CSI, the performance of beam searching is closed to its upper bound when the sampling rate of the active element is larger than $1/8$.

\subsection{Beam Tracking in High Mobility Scenario}
For high mobility scenario, the main challenge of the RIS beamforming design is to precisely match the time-varying characteristics.
In \cite{Drones}, the authors consider a THz drone network, where a mobile user is served by a BS and a flying RIS.
Based on the prior observations of the beam trajectories, a DL algorithm is proposed to proactively predict the optimal RIS beamforming.
Due to the mobility of the drone, a Gated Recurrent Unit (GRU) based network structure is utilized for capturing the temporal correlations  within the spectrum data and learning the sequence dependency.

Besides, motivated by the  ODE structure in \cite{Meng_Xu_RIS} and LSTM framework, we introduce another feasible scheme integrating the tracking of both the channel and the beam.
First, the RIS assisted channel can be sub-sampled through closing/activating partial RIS elements.
Similar to the LSTM based time domain channel extrapolation, one may construct an ODE based time-varying channel model, acquire the sub-sampled channel over the time domain, and then implement the channel tracking through LSTM.
With the acquired antenna domain sub-sampled channel at any expected time, the optimal beamforming matrix at the corresponding time can be selected through the classification network \cite{Shunbo_extrapolation}.

\section{AI Assisted Resource Scheduling}

The optimization objectives of the AI assisted resource scheduling cover the following aspects: the spectrum and energy efficiency, the environmental sensing range,  the cost of system deployment, and the network topology, etc.
On the other hand, the feasible operations for the resource scheduling include the decision for the on/off status of the RISs, the joint beamforming design of the multiple BSs and RISs, and the joint power allocation.
Besides, in high mobility scenarios, since the channel links are strongly related to the schedules on the RISs and the BSs, the user mobility will further affect the optimal network topology of the RIS communications.
Thus, the network topology optimization in the mobility scenario should be discussed.
The scenario of RIS communications and the concept of AI based resource scheduling is presented in \figurename{ \ref{fig:4}}.

\subsection{RIS Scheduling for Serving Range Enlargement}
\begin{figure}[!t]
 \centering
 \includegraphics[width=3.3in]{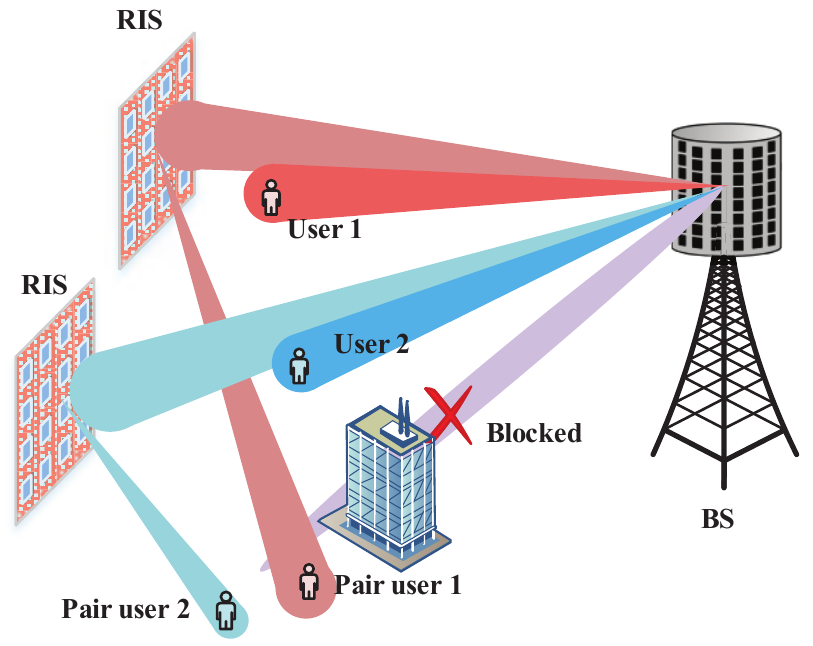}
 \caption{Illustration of user partitioning and RIS beamforming for RIS based NOMA scenario.}
 \label{fig:NOMA}
\end{figure}

To maximize the sum rate of the downlink non-orthogonal multiple access (NOMA) networks,
the authors in \cite{RIS_NOMA1} invoke a modified object migration automation algorithm to partition the users into several equal-size clusters, which is shown in \figurename{ \ref{fig:NOMA}}.
Then, they propose a deep deterministic policy gradient (DDPG) algorithm to collaboratively control multiple reflecting elements at RIS.
Different from the training-then-testing process, the authors adopt a long-term self-adjusting learning scheme to search  the optimal action in every given environment, i.e., the user partitioning and RIS beamforming, through iterative exploration-and-exploitation.
This RIS aided NOMA downlink network achieves higher sum data rate than the conventional orthogonal multiple access (OMA) one.
Besides, the DDPG algorithm can effectively learn the dynamics of the resource allocation policies in a long-term manner.

To improve the performance of RIS assisted mmWave communications, the authors of \cite{mmwave11} propose a  machine learning empowered beam management  framework to precisely align  beams  between BSs and the user terminals.
Within this framework, a DNN is utilized to recognize the complicated network environment, which can help each BS select  its optimal RIS.
Besides, an online learning method is adopted to predict the mobility of the terminals.
By integrating environmental and mobility awareness, the proactive  beam-switching scheme can be achieved by mapping the user's position to its optimal RIS.

In addition to many works with a single RIS, there are also many works discussing the cooperation of multiple RIS.
For example,
the channel links in RIS communications can also be scheduled in a multi-hop way, which can effectively combat the severe propagation attenuations and improve the coverage range of the environmental sensing.
However, due to the multiuser interference, mathematically intractable multi-hop signals, and the non-linear constraints, the formulation of the optimization problem on maximizing  the sum rate of the multi-RIS system is non-convex.
In \cite{THz_DRL}, the authors propose a DRL based algorithm to enlarge the sensing range of the RIS communications with the DDPG algorithm.
To overcome the impractical search complexity of the extremely large Q-table in the DRL framework, a DNN based Q-learning method is utilized.
Besides, the hardcore of DDPG structure is the critic and actor network that is  comprised of a fully connected DNN.
Inputting the beamforming matrix and the channel information, the DDPG algorithm will output the beamforming designs for the current iteration.
It is shown in \cite{THz_DRL} that the covering range can be improved more than 50\% with the DRL based scheme.

\subsection{RISs and BSs Scheduling in the Static Scenario}
In \cite{Scheduling_DRL}, the authors adopt the DRL approach and maximized the average energy efficiency by jointly optimizing the transmit power allocation, RIS beamforming matrix, and RIS elements' ON/OFF state.
Although the predefined reward may be affected by various unknown aspects such as the uncertainty on the channel of the users,
the difference between the expected reward and the actual reward can be used by the DNN in the DRL framework with the feedback returned from the uncertain environment.

While the resource scheduling in RIS communications involves the training of the NNs for a number of users,
the FL framework is also a feasible approach to deal with the resource scheduling problem.
In \cite{mmwave_FL}, the authors
apply FL for RIS assisted mmWave system and  achieve  a privacy-preserving design.
The FL is applied to train the  DNN model for the mapping between user's channel and its optimal RIS beamforming.
Moreover,  FL can help optimize multiple RISs in parallel while protect private CSI.

However, since all local parameters are transmitted over shared wireless channels, the undesirable propagation error will inevitably deteriorates the performance of global aggregation.
In \cite{FL_multiRIS}, the authors propose an alternating optimization algorithm to minimize the aggregation error and accelerating the convergence rate of federated learning.
By invoking semi-definite relaxation methods, the non-linear and non-convex problem for jointly optimizing the transmit power, the receive scalar, and the phase shifts is solved.
It is worth noting that leveraging distributed RISs to reconfigure the wireless channels can significantly improve the performance of this scheme.

In addition, the location information of the BSs, the RISs, and the users can also assist the resource scheduling.
With the knowledge of the location information, the location relationship between different objects, i.e., the BS and the RIS, or the RIS and the users, can be easily obtained through geometrical methods.
With a predefined set for beamforming selection, a properly designed NN can be utilized to search the mapping between the location information and the optimal beamforming index.

\subsection{RISs and BSs Scheduling for Mobility Scenario}
In mobility scenario, since the positions of the users are changing, the adaptation of the varying environment is necessary for the resource scheduling.
In \cite{Scheduling_Timevarying_Multicell}, to maximize the uplink sum-rate of the multi-RIS assisted multi-cell network in mobility scenario, the authors propose a dynamic control scheme based on the multi-agent DRL, in which the BSs are regarded as independent agents.
In particular, each BS adaptively  configure its local UE powers, local RIS beamformer, and its combiners.
For non-stationarity caused by the coupling of multiple BSs' actions, an efficient message passing scheme is introduced, which requires limited information exchange among neighboring BSs.
Compared with the maximum ratio combining method,  \cite{Scheduling_Timevarying_Multicell} performs better in terms of the average data rate.

\begin{figure}[!t]
 \centering
 \includegraphics[width=3.3in]{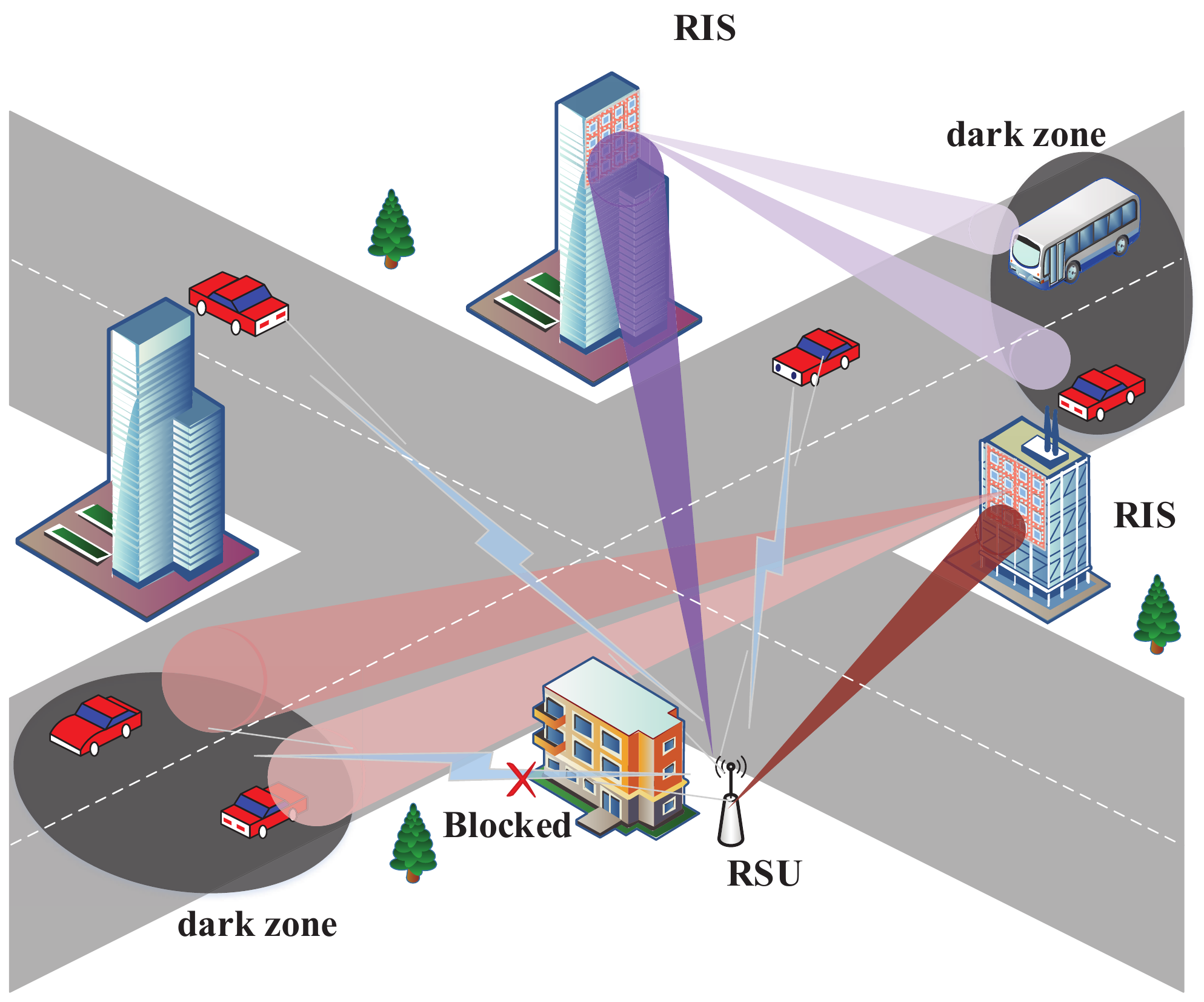}
 \caption{RIS scheduling in vehicular communication scenario.}
 \label{fig:VC}
\end{figure}

Especially, for the vehicular communications, the RoadSide Unit (RSU) can leverage RIS to provide indirect wireless transmissions within the uncovered areas (called the ``dark zones''), which is presented in \figurename{ \ref{fig:VC}}.
However, the vehicles move at different directions and  mobility speeds, and
have various residence time in the dark zones.
If same traffic is offered to all the vehicles within the dark zones,
the low mobility vehicles will achieve poor service quality than the high ones.
Hence, in \cite{vehicle_comm}, the joint vehicle scheduling and passive beamforming in RIS empowered vehicular communication are investigated to maximize the minimum achievable rate for the vehicles in the dark zone.
Specially, the whole problem is decoupled into two sub-ones: the wireless scheduling and the RIS phase-shift optimization.
The former one is solved through DRL to obtain the optimal RSU scheduling schemes under various road conditions and RIS options, while the later one is settled by the block coordinate descent (BCD) algorithm to maximize the instantaneous sum rates for all the currently served vehicles.
The BCD algorithm is demonstrated to have a great robustness.

\subsection{Improvement of the DRL-based Resource Scheduling Methods}
It is worth noting that one common characteristic of previously introduced DRL methods is the requirement of CSI as the input of the learning structures.
However, due to the unpredictable hardware impairment or the utilization of hybrid passive/active RIS, only partial CSI of the RIS communications may be acquired.
One alternative approach is to first extrapolate the full CSI from the partial one with the aforementioned channel extrapolation schemes, and regards the acquired full CSI as the input of the DRL frameworks.
Nevertheless, since the scale of the channel in RIS communication is very large, the whole computation complexity of the DRL for resource scheduling will be unaffordable.
Motivated by the beam searching scheme in \cite{Shunbo_extrapolation}, one can
add a classification network in the front of the existed DRL frameworks to learn the mapping between the partial CSI and the optimal resource scheduling strategies, including the BSs power allocation, the RISs beamforming, etc., and then further assist the actions choosing operation of the DRL.

\section{Future Research Directions}
Although some theoretical and implemental AIRIS strategies  have been proposed, there are still many open issues to further enhance the intelligent usage of RIS. We here highlight several potential directions below for continuous research on AIRIS.

\subsection{The AI assisted Integration of Sensing and Communication}
AI can also facilitate the integration of communication, imaging, sensing, and localization.
For example, with the knowledge of the environmental image, one can recognize the location information of the blocks and the users, and directly select the feasible RISs for constructing the optimal communication links for the BS and one specific user  \cite{Gao_image_beam}.
On the other hand, the acquired channel information can also be utilized to extract the feature of the surrounding environments.
Notice that the precise sensing results can also be achieved through obtaining partial CSI with low cost, as did in Section 5.4.
With the extrapolation, one can design an NN to approximate the mapping between the partial CSI and the high-precision imaging and localizations.
Hence, a AI based robust solution can be designed to deal with the two problems in turn.

\subsection{AI based Beamforming Design for Wideband RIS communications}
The existing works mainly focus on the narrow band communications.
However, in the wideband case, the channel matrices of the different sub-carriers possess distinct subspaces, which is called the beam squint effect \cite{Wideband_gao}.
Hence, traditional hybrid precoding schemes will suffer from an eroded performance \cite{Wideband2}.
Inspired by \cite{Wideband1}, we can adopt the non-uniform quantization codebook based RIS beamforming.
Similar to \cite{Shunbo_extrapolation}, a classification network based beam searching can be utilized to optimize the RIS's operation.
To alleviate the effect of beam squint, a DL based phase compensation operation can be implemented after beam searching.

\subsection{Transfer Learning for Model Generalizations}
In this article, we have introduced several  DRL based  environment adaptation strategies for RIS communications, where the model training of different users are implemented via either the centralized way or federated way.
While the FL can help reduce the transmission overhead of the training dataset, the update of the model parameters for different users are still individually implemented.
For multiple users in a similar environment, since AI can exploit the intrinsic relevance between the channel measurement or parameters of different users, the transfer learning can be applied to accelerate the training process. Specifically, when the NN of one user was trained, its inner parameters can be transferred into other models, and further generalize the trained NN to a similar one  \cite{YuwenYang_transfer_learning}.

\subsection{AI based RIS Pre-Deployment}
To cover as many communication scenarios as possible, an appropriately over-saturated RIS pre-deployment should be implemented \cite{RISdeployment_large}.
Since RISs are always placed on the outside wall of the buildings for the outdoor scenarios or on the wall of rooms for the indoor scenarios, there are finite alternative positions for the deployment of RISs in a certain environment.
Hence,   RIS pre-deployment  can be regarded as a pattern selection problem, and a selection network can be utilized to maximize the achievable rate of the system.
Besides, the characteristics of activating users and the RIS deployment costs should also be considered in  the loss function.
Since the pre-deployment does not have a very high real-time requirement, the long training time is acceptable.

In addition, RISs can also be deployed at the aero-crafts, i.e., the unmanned aerial vehicles (UAVs) \cite{UAV_RIS}.
Under this situation, the BS-RIS link become dynamical and will bring dual-time-varying channel models.
New dimensions, like the UAV's position and trajectory, can be designed to sustain the optimal network topology for  the whole system.

\section{Conclusion}
In this paper, we introduced the recent researches on the integration of AI and RIS communications, i.e., AIRIS.
Compared to traditional signal processing techniques,  AI can achieve a higher degree in all aspects of RIS communications, such as the environmental sensing, channel acquisition, the beamforming design, and the resource scheduling.
A number of neural network structures were introduced, such as the CNN, the ODE inspired LSTM, the RNN, and the DNN.
Besides, we also provide several potential directions for future AIRIS communications, i.e., the integration of sensing and communication, the transfer learning for model generalizations, and the AI based RIS pre-deployments design. It can be expected that the AIRIS will significantly facilitate the robustness of the transmission, and will play an important role in the realization of B5G and 6G.

\balance

%


\end{document}